# Dynamics of Gender Bias in Software Engineering

Thomas J. Misa

Abstract: The field of software engineering is embedded in both engineering and computer science, and may embody gender biases endemic to both. This paper surveys software engineering's origins and its long-running attention to engineering professionalism, profiling five leaders; it then examines the field's recent attention to gender issues and gender bias. It next quantitatively analyzes women's participation as research authors in the field's leading International Conference of Software Engineering (1976-2010), finding a dozen years with statistically significant gender exclusion. Policy dimensions of research on gender bias in computing are suggested.

Software engineering is not yet a true engineering discipline, but it has the potential to become one. — Mary Shaw (1990)[1]

From the moment of its formal origin in 1968, software engineering has repeatedly aimed to identify and corral a domain of professional engineering practice.[2] Software engineering is among the numerous computing disciplines, including computer science and allied fields, that have in recent years grown massively to constitute nearly half of the entire US STEM workforce.[3] Yet, alarmingly, computing as a whole displays persistent and worrisome patterns of gender bias in education, training, recruitment, and workforce participation. Software engineering may be a classic instance of engineering with male-dominated leadership, culture, and normative mechanisms of reproduction, as *Engineering Studies* authors have repeatedly suggested.[4]  Still, software engineering appears distinctive in the frequency that its leaders invoke stylized models of engineering professionalization, such as Mary Shaw's epigram above.[5]

---

[1] Mary Shaw, "Prospects for an Engineering Discipline of Software," *IEEE Software* 7, no. 6 (1990): quote p. 15.

[2] See e.g. the four ICSE papers introduced by James E. Tomayko, "Twenty-year Retrospective: The NATO Software Engineering Conferences," *11th International Conference on Software Engineering* (Pittsburgh, PA, 1989): 96-96 at doi.org/10.1109/ICSE.1989.714397.

[3] In 2019, the US Bureau of Labor Statistics reported that of 9.9 million STEM jobs, 4.6 million (or 46 percent) were in computer occupations, with the remainder comprised by engineers, life sciences, STEM post-secondary teachers, physical sciences, and mathematical sciences (in descending order): Alan Zilberman and Lindsey Ice, "Why Computer Occupations are Behind Strong STEM Employment Growth in the 2019-29 Decade," *Beyond the Numbers* (January 2021) at https://www.bls.gov/opub/btn/volume-10/why-computer-occupations-are-behind-strong-stem-employment-growth.htm

Gender and professionalization is a complex and timely topic.  Recently, software engineers themselves have examined gender bias in their field[6]; another sign of professional awareness is a set of five international conferences on gender equity in software engineering (2018 et seq.).  The present article contributes a new longitudinal dataset (N=6,000) analyzing women's research contributions to the preeminent International Conference on Software Engineering (1976–2010).

This article proceeds as follows.  It first outlines the history of software engineering and its practitioners' professionalizing strategies.  Prominent figures including Doug McIlroy, Barry Boehm, Mary Shaw, Niklaus Wirth, and others have explicitly mobilized professionalizing discourse.  The article next summarizes practitioners' recent attention to gender bias, including five international workshops on gender equity in software engineering, as well as reviews gender bias in the wider field of computing.  The following sections present new longitudinal data on women's changing participation as research authors in software engineering.  Three distinct stages, with different gender dynamics, can be discerned across 35 years.  Finally, the article's conclusion suggests the policy salience of research on gender bias in software engineering.

---

[4] See Vivian Anette Lagesen and Knut H. Sørensen, "Walking the Line? The Enactment of the Social/Technical Binary in Software Engineering," *Engineering Studies* 1 no. 2 (2009): 129-149; Pedro Sanches, "Seeing Mobility: How Software Engineers Produce Unequal Representations," *Engineering Studies* 8 no. 1 (2016): 27-47; and Kacey Beddoes, "Taking Stock and Looking Forward: Fifteen Years of Research on Gender, Race, and Power in *Engineering Studies*," *Engineering Studies* 16 no. 1 (2024): 1-7.

[5] Besides the five visions discussed below, see R. L. Baber, "Comparison of Electrical 'Engineering' of Heaviside's Times and Software 'Engineering' of Our Times," *IEEE Annals of the History of Computing* 19, no. 4 (1997): 5-17 at doi.org/10.1109/85.627895

[6] Christoph Treude and Hideaki Hata, "She Elicits Requirements and He Tests: Software Engineering Gender Bias in Large Language Models," ArXiv (2023) at doi.org/10.48550/arXiv.2303.10131; Emitzá Guzmán, Ricarda Anna-Lena Fischer, and Janey Kok, "Mind the Gap: Gender, Micro-Inequities and Barriers in Software Development," *Empirical Software Engineering* 29 no. 17 (2024) n.p. at doi.org/10.1007/s10664-023-10379-8 ; Sarah Alkoblan, Hala Almukhalfi, Sahar Alturki, Muna Al-Razgan, "Gender Diversity in Software Engineering: A Systematic Mapping," *International Journal of Computer Applications* 186 no. 46 (2024): 24-37 at doi.org/10.5120/ijca2024924098 ; Letizia Jaccheri and Anh Nguyen Duc, "Software Engineering and Gender: A Tutorial," In *Companion Proceedings of the 32nd ACM International Conference on the Foundations of Software Engineering* (New York: ACM, 2024): 704-706 at doi.org/10.1145/3663529.3663818

## Origins of software engineering

Its practitioners most frequently trace the origins of "software engineering" to an international conference organized by the Science Committee of the North Atlantic Treaty Organization (NATO) that brought 50 experts to the mountain town of Garmisch in southern Germany, for five days in early October 1968. Attendees included academic researchers, computer manufacturers, software developers, and computer users; the term software engineering "was deliberately chosen as being provocative." Software engineering implied "the need for software manufacture to be based on the types of theoretical foundations and practical disciplines, that are traditional in the established branches of engineering."[7] According to historians Janet Abbate and Nathan Ensmenger the NATO conference was a response to an ill-defined "software crisis" of the 1960s as managers of large programming projects sought to impose quantitative methods, discipline the programming workforce, and adopt high-status "engineering" models and approaches.[8]

An alternative to the NATO origin story circulates around the figure of Margaret Hamilton, widely recognized in the NASA–Apollo mission's software development. Hamilton, at some time (alas the chronology is not quite clear), promoted the term "software engineering" in her work at MIT's (Draper) Instrumentation Laboratory, which she joined in 1965 after three years at MIT's Lincoln Laboratories working on the massive air-defense system Project SAGE. A positive case is ready at hand that she "coined the term software engineering"; is "famous for building Apollo's on-board flight software and inventing the term 'Software Engineering' to establish it as a form of engineering in its own right"; and creatively "made up" the term to gain legitimacy for the software field. In a 2018 essay, she herself wrote "Little did we know that what we were doing then [at MIT circa 1960] would become known years later as 'software engineering,' when we were in the trenches building flight software for the Apollo missions."9 Hamilton's work may have been especially influential during the years between the NATO conference (1968) and the first International Conference on Software Engineering (1976), discussed below.

---

[7] Peter Naur and Brian Randell, eds., *Software Engineering: Report of a Conference Sponsored by the NATO Science Committee* (Brussels: NATO Science Committee, 1969), quote p. 13. The text is most commonly found at http://homepages.cs.ncl.ac.uk/brian.randell/NATO/nato1968.PDF

[8] Janet Abbate, *Recoding Gender: Women's Changing Participation in Computing* (MIT 2012), 73-75, 98-111; Nathan Ensmenger, *The Computer Boys Take Over: Computers, Programmers, and the Politics of Technical Expertise* (MIT 2010), 195-204, 218-21. Conversely, Thomas Haigh is decidedly skeptical about any such link in his "Crisis, What Crisis? Reconsidering the Software Crisis of the 1960s and the Origins of Software Engineering" (2010) at https://tomandmaria.com/tom2/Writing/SoftwareCrisis_SofiaDRAFT.pdf

It is worth emphasizing that computing in these years was rapidly expanding and massively in transition. "Software" was scarcely a decade old as a defined term describing computer programs, often written by large computer manufacturers like IBM and Univac or government contractors such as RAND and Systems Development Corporation, with the first US software patent issued in 1968. Accordingly, the NATO conference's notion of "software manufacture" along mass production or factory lines was wholly aspirational and future-looking.10 Computer science was just emerging as a well-defined professional field; the Association for Computing Machinery (ACM) was developing model university curricula; the National Science Foundation's discipline-building role as well as the National Academy of Engineering's model curricula for computer engineering was yet to come.

Compared with the classic engineering professionalism of the late nineteenth century, software engineering in 1968 had surprisingly substantial industry participation. Garmisch was not a gathering of independent, high-status, gentleman engineers like Alexander Holley and Robert Thurston of the ASME. Large companies like IBM, Bell Labs, General Electric, Mobil, Scientific Data Systems, Control Data, Siemens, and Philips sent the most representatives (49 percent), larger than government installations and agencies (16 percent); with academics totaling most of the rest (31 percent). Independent consultants were rare (just two).[11] Brian Randell (from IBM's Watson Research Center), who co-edited the conference volume with Peter Naur (from Denmark's pioneering computer company Regnecentralen), articulated five conference themes: software as commodity, programming languages, multiprogramming and time-sharing, modularity and structuring, and problems of large systems.[12] Each merits brief elaboration, to make clear how software engineering was effectively created around distinctive industry-scale problems.

---

9 See, respectively, Lori Cameron, "Margaret Hamilton: First Software Engineer," *IEEE Computer Society Tech News* (5 October 2018); Message from the ICSE 2018 General Chair, quote p. vii; and "Margaret Hamilton," 2017 Computer History Museum Fellow [citation] at computerhistory.org/profile/margaret-hamilton/ (accessed August 2024). See page 33 of her 2017 interview with David C. Brock at https://archive.computerhistory.org/resources/access/text/2022/03/102738243-05-01-acc.pdf; also her keynote address to ICSE 2018 "The Language as a Software Engineer" (video at https://www.youtube.com/watch?v=ZbVOF0Uk5lU ). Hamilton, "What the Errors Tell Us," *IEEE Software* 35, no. 5 (2018): 32-37, quote 32, at doi.org/10.1109/MS.2018.290110447

10 In 1968 Anthony Oettinger described software engineering in aspirational and idealistic terms: "it's a shining ideal — the software engineer … walks with the Gods as well as with the lowliest of creatures." See his "The Role of the Computer Programmer in Society," In *Proceedings of the Sixth SIGCPR Conference on Computer Personnel Research* (New York: ACM, 1968): 4-10 at doi.org/10.1145/1142648.1142650. More measured is Bernard A. Galler, "NATO and Software Engineering?" *Communications of the ACM* 12, no. 6 (1969): 301 at doi.org/10.1145/363011.363013

Aspirations to create *software as a commodity* like standardized wheat or coal ran square into the central market position of IBM, which with six attendees was by far the most visible computer company at Garmish.[13] For its landmark IBM model 360, launched four years earlier, IBM intentionally bundled its software and hardware into a single package, often tailored to specific computer users and quite distinct from a market commodity. The tantalizing prospect of IBM's unbundling its software and hardware was a "hot issue" for the NATO conference attendees, hinting at an independent software industry. Industry participant Robert Bemer advocated a full-blown "machine-controlled production environment, or software factory." IBM, already exploring similar lines, had found to its dismay: "The cost is enormous, and a vast amount of hardware is needed."[14] Commodities did not come cheap.

While academic researchers actively analyzed new *programming languages*, such as Fortran, COBOL, and ALGOL, the additional three NATO conference themes were particularly relevant to large organizations in industry and government. Giant mainframe computer installations were the chief loci for *multiprogramming and time-sharing*, where more than one program, or more than one program user, respectively, could simultaneously access a single computer's hardware.  In 1968 there were around 70 installations worldwide of 30 distinct computer systems offering such access.  Some advocated the intentional use and reuse of code segments, procedures, blocks, and subroutines; in this way, software engineering could impose order on computer programs, sometimes dubbed spaghetti code, by creating programming environments to discipline and structure the work of computer programmers.  Noted one participant, "one can find little mention of these ideas elsewhere in the academic computing literature in 1968."[15] Complex code pointedly raised the *problems of large systems*, such as American Airlines' SABRE reservation system with 300,000 instructions and IBM's 360 operating system with a million instructions.  Prospects for a vastly intricate computer-centric anti-ballistic missile system soon brought these concerns to widespread public attention.[16]

---

[11] See Naur and Randell, *Software Engineering*, pp. 214-217

[12] Brian Randell, "Software Engineering in 1968," in *Proceedings of the 4th International Conference on Software Engineering* (IEEE Press, 1979): 1-10.

[13] Bell Telephone Laboratories sent three, with the British Ministry of Technology, MIT, and University of Edinburgh each sending two representatives to Garmish. On Bell's profile in digital computing, see Kim W. Tracy, "Bell Labs's Portrayal of Switching as Computing (or Not)," *IEEE Annals of the History of Computing* 46 no. 3 (2024): 81-85 at doi.org/10.1109/MAHC.2024.3437008

[14] Robert Bemer had worked at RAND, Lockheed, IBM, Sperry Rand Univac, General Electric and after 1968 also Honeywell. See "Robert W. Bemer," in J.A.N. Lee's *Computer Pioneers* (Los Alamitos: IEEE Computer Society Press, 1995) online at https://history.computer.org/pioneers/bemer.html. IBM's Ascher Opfer quoted at https://web.archive.org/web/20010406041743/http://www.bobbemer.com/FACTORY.HTM

In imagining a future for software engineering, Garmish participants invoked the engineering of bridges, automobiles, airplanes, and even the industrial revolution.  As Bell Laboratories' Doug McIlroy, one of the conference organizers, vividly phrased his vision for software engineering, "the software industry is not industrialized."  In his plenary address, much quoted ever since, he stated:

> We undoubtedly produce software by backward techniques. We undoubtedly get the short end of the stick in confrontations with the hardware people because they are the industrialists and we are the [pre-industrial] crofters. Software production today appears in the scale of industrialization somewhere below the more backward construction industries. I think its place is considerably higher, and would like to investigate the prospects for mass-production techniques in software.[17]

---

[15] Randell 1979, quote p. 3

[16] C. R. Vick, "First Generation Software Engineering System: The Ballistic Missile Defense Software Development System." In *Proceedings of the 1977 Annual Conference* (New York: ACM, 1977): 108-114 at doi.org/10.1145/800179.810190 ; Rebecca Slayton, *Arguments that Count: Physics, Computing, and Missile Defense, 1949-2012* (MIT Press, 2013).

[17] McIlroy later observed "software was not organized. Engineering was, and software did not really play a role in the design of computers. Software was something that you left [for later]." See p. 53 and quote p. 59 of David C. Brock, Oral History of Malcolm Douglas (Doug) McIlroy, Computer History Museum X9183.2020 (September 30, 2019) at https://archive.computerhistory.org/resources/access/text/2019/10/102795421-05-01-acc.pdf

The industrial revolution and mass-production technologies then enjoyed substantial prestige around the world as McIlroy clearly understood and shrewdly mobilized.[18] These professional aspirations cast a long shadow, as a review of four additional visions beyond McIlroy's—advanced by Barry Boehm, Mary Shaw, Nancy Leveson, and Niklaus Wirth—attest.

An early and forceful advocate of software engineering, Barry Boehm learned about the NATO conference from a RAND colleague who had attended. The large-scale emphasis of software engineering spoke directly to Boehm.[19] His notable career at RAND, the Air Force, and TRW, a leading military contractor, placed him at the center of high-level efforts to design and develop computer-guided missile and anti-ballistic missile defense systems (exactly the large systems noted above). As Boehm recalled "I was captivated by the prospect of making software engineering into a quantitative discipline." Becoming TRW's Director of Software Research and Technology in 1973, "I thought that a fully quantified software engineering methodology would be complex and take a good five years to work out."[20]

Boehm also personally connected the 1968 NATO conference with the first International Conference on Software Engineering in 1976. In Europe, he met with NATO conference organizers and luminaries such as Friedrich Bauer (conference chair), Brian Randell (noted above), Edsger Dijkstra, and Tony Hoare. An advocate of structured programming, Dijkstra in 1968 published the polemical "Go To Statement Considered Harmful" and soon helped Boehm refine his ideas on software reliability. With Hoare, Boehm was program co-chair for the 1975 joint ACM–IEEE Conference on Software Reliability, which spawned the 1976 International Conference on Software Engineering, or ICSE, the first in a long-running series (analyzed below). As noted above, Margaret Hamilton's influence on software engineering following the Apollo moon landing in 1969 might be decisive for sustaining the NATO term and so (indirectly) creating the ICSE.[21]

---

[18] Michael Mahoney suggested varied "models" used by software engineering (science, engineering, industry, and professions), noting that the "software factory" offered 20th century models of Fordist mass production. See Mahoney's insightful "Finding a History for Software Engineering," *IEEE Annals of the History of Computing* 26 no. 1 (2004): 8-19; Michael A. Cusumano, *Japan's Software Factories: A Challenge to U.S. Management* (Oxford University Press, 1991), notes that SDC in the mid-1970s adopted a "factory" model for software development (putting ten percent of its programming workforce in a special purpose-built facility) but that within three years apparently abandoned it as unworkable and overly complex.

[19] Boehm, "Software Engineering," *IEEE Transactions on Computers* C-25, no. 12 (1976): 1226-1241 at doi.org/10.1109/TC.1976.1674590 ; Richard W. Selby, "Software Engineering: The Legacy of Barry W. Boehm," *29th International Conference on Software Engineering* (Minneapolis, MN, 2007): 37-38 at doi.org/10.1109/ICSECOMPANION.2007.67.

[20] Hakan Erdogmus and Nenad Medvidovic, "A Conversation with Barry Boehm," *IEEE Software* 35, no. 5 (2018): 14-19, quotes p. 16, at doi.org/10.1109/MS.2018.3571249. Later, Boehm was Director of the Defense Advanced Research Projects Agency's Information Science and Technology Office and Director of the Defense Research and Engineering's Software and Computer Technology Office from 1989 to 1992.

Mary Shaw earned her PhD in computer science from Carnegie Mellon University in 1972, immediately joined its faculty, and played a key role in its pioneering Software Engineering Institute and serving as its chief scientist (1984-87).[22] Her distinguished career brought both ACM's and IEEE's recognition of her achievements in software engineering, culminating with the US National Medal of Technology and Innovation (2014). Her 1990 article in *IEEE Software*, entitled "Prospects for an Engineering Discipline of Software" (quoted in the epigram) articulated a forward-looking view for how the nascent field could make progress from a pre-industrial "craft" (the crofters above) through two schematic stages, to become a full "professional engineering" discipline.[23] To begin, "production" and "craft" were fused into a first stage labeled "commercial"; and in the second stage "science" could help bring about "professional engineering." To elaborate her schema, Shaw offered two historical mini-studies. Civil engineering achieved the first stage with Roman engineering in the first century CE. It later drew on scientific results in statics and strength of materials emerging around 1700 to achieve the second stage with engineering properties of materials (1750) and the complete engineering analysis of bridges (1850) such as Robert Stephenson's Britannia Bridge in Wales. Chemical engineering achieved the first stage with the Leblanc alkali process (1790), then, owing something to atomic theory (1800), the advent of chemical engineering's "unit operations" (1890-1915) brought about the second stage. "Chemical engineering as a science … is not a composite of chemistry and mechanical and civil engineering, but a science of itself," noted one observer.

---

[21] Patricia A. Hamilton (1978) and David J. Hamilton (2010) were authors of ICSE papers 1976-2010 but Margaret Hamilton apparently was not.

[22] Mary Shaw, "Remembrances of a Graduate Student," 11th International Conference on Software Engineering (Pittsburgh, PA, 1989): 99-100 at doi.org/10.1109/ICSE.1989.714401.

[23] Mary Shaw, "Prospects for an Engineering Discipline of Software," *IEEE Software* 7 no. 6 (1990): 15-24 at doi.org/10.1109/52.60586. A follow-on was Mary Shaw, "Continuing Prospects for an Engineering Discipline of Software," *IEEE Software* 26 no. 6 (2009): 64-67 at doi.org/10.1109/MS.2009.172.

Shaw then boldly charted a professional path for software engineering. Its first stage had recently been achieved with specific software-development methodologies in the 1980s. Computer science research in algorithms and data structures pointed to the second phase, but Shaw admitted that only "isolated examples" existed, such as compiler construction. She forecast an engineering discipline of software could result from short-term attention to "pragmatic, possible purely empirical contributions that help stabilize commercial practice" as well as longer-term efforts "to develop and make available basic scientific contributions."

A stylized industrial revolution again took center stage in Nancy Leveson's 1992 ICSE keynote address "High Pressure Steam Engines and Computer Software." With a PhD from UCLA, Leveson was at the time a faculty member at University of California–Irvine, soon to be Boeing Professor of Computer Science and Engineering at University of Washington, and then later a professor in MIT's Aeronautics and Astronautics Department, working on software engineering and safety-critical systems.[24] Leveson found a promising parallel between contemporaneous software engineering and mechanical engineering in the late nineteenth century. Her touchstone was the promise of engineering professionalism, based on ASME's founding president Robert Henry Thurston's classic *History of the Growth of the Steam-Engine* (1878).[25] This was not such a stretch. ASME even today showcases its longterm professional concern with "the safety of equipment used in manufacturing and construction, particularly [steam] boilers and pressure vessels," based on an influential suite of engineering standards (1884-1915).[26]

---

[24] Nancy G. Leveson, "High-Pressure Steam Engines and Computer Software," in *Proceedings of the 14th International Conference on Software Engineering* (New York: ACM, 1992): 2-14 at doi.org/10.1145/143062.143076. For Leveson's bio see https://archive.ph/STpCU . With Maria Klawe, Leveson in 1991 co-founded the Computing Research Association's Committee on the Status of Women in Computing Research, or CRA-W.

[25] Thurston's *History* was published in New York by Appleton in 1878; ASME was founded two years later; Leveson consulted the 1883 edition published in London.

[26] See ASME's "Engineering History" at https://web.archive.org/web/20240401031520/https://www.asme.org/about-asme/engineering-history ; JoAnne Yates and Craig N. Murphy, *Engineering Rules: Global Standard Setting since 1880* (Baltimore: Johns Hopkins University Press, 2021), 31-46.

Specifically citing ASME's uniform boiler codes for managing the risks of exploding steam boilers, Leveson asked, provocatively, "Exploding Software?" There was some cause for concern. She pointedly contrasted the relatively simple and even verifiable 6,000 lines of software used by Ontario Hydro for one computerized nuclear shutdown system with the overly complex 100,000 lines used in England's Sizewell B nuclear reactor for both control and shutdown that was "beyond our ability to apply sophisticated software verification techniques [and] violates the basic nuclear reactor safety design principle that requires complete independence of control and safety devices." With the ill-fated Therac-25, an unwise over-reliance on software, "when designers of this radiation therapy machine eliminated the usual hardware safety interlocks," resulted in four deaths from massive radiation overdoses (1985-87).[27] "Like the exploding boilers," she concluded on a positive professionalizing note: "our ability to build safe software-controlled systems and to build effective software engineering tools to accomplish this will be enhanced by greater understanding of the scientific foundations of our craft." (8)

Professionalism in engineering often mobilizes its history. Indeed, in his 2008 "Brief History of Software Engineering" Niklaus Wirth (1934-2024) gave a polite bow to the NATO conference which, he said, launched software engineering as "the highly disciplined, systematic approach to software development and maintenance" especially of complex computer systems.[28] After completing his Ph.D. at University of California–Berkeley, Wirth had worked closely with Dijkstra and Hoare on the ALGOL project in the 1960s, then developed his influential language Pascal, widely used in teaching computer science and readily adapted to the small memories of microcomputers in the 1970s. Wirth was an early advocate of structured programming and modularization, "the most important contribution to software engineering" (36). He believed that software engineering was trapped between excessively structured languages such as the Defense-Department's Ada (which needed five lines of code for the simple "Hello World"[29] test) and other popular languages spreading in the programming community such as Unix, "C," and open source that permitted programmers to subvert the disciplining structures.[30] "The trouble was that C's rules could easily be broken, exactly what many programmers valued" but which resulted in "pitfalls that made large systems error-prone, and costly to debug and 'maintain'." (p. 35). "From the point of view of software engineering [this] represented a great leap backward." (34). Wirth, professor of computer science at the Swiss Federal Institute of Technology (known as ETH Zurich) for decades, focused on education as a way to achieve professional stature.[31] "Software engineering would be the primary beneficiary of a professional education in disciplined programming." (39)

---

[27] See Nancy G. Leveson and Clark S. Turner, "An Investigation of the Therac-25 Accidents," *IEEE Computer* 26, no. 7 (1993): 18-41 at doi.org/10.1109/MC.1993.274940; and Leveson, "The Therac-25: 30 Years Later," *IEEE Computer* 50, no. 11 (2017): 8-11 at doi.org/10.1109/MC.2017.4041349.

[28] Niklaus Wirth, "A Brief History of Software Engineering," *IEEE Annals of the History of Computing* 30 no. 3 (2008): 32–39 at doi.org/10.1109/MAHC.2008.33. See also Grady Booch, "The History of Software Engineering," *IEEE Software* 35 no. 5 (2018): 108-114 at doi.org/10.1109/MS.2018.3571234.

**Dynamics of gender bias**

The above section profiled five authors of influential software-engineering visions by name; intentionally, two were women. In the mid-1980s women constituted roughly two in five members of the high-skilled computing workforce in the United States and collected nearly two in five undergraduate computer science degrees. In the US, computing from the 1960s to the mid-1980s was distinctively attractive to women especially compared with the physical sciences and other branches of engineering. Unfortunately, since the mid-1980s these proportions have fallen so that in recent years women are roughly *one in five* in both workforce and undergrad computer-science degrees, varying in different national contexts. There is wide agreement that these numbers are too low, clearly blocking women's full participation in the well-paying computing workforce, inevitably reducing desirable diversity in computing research and development efforts, and possibly leading to undesirable biases in computing devices and systems that are central to modern society.[32]  This section briefly reviews research on gender bias in computing by analyzing five recent international conferences on gender equality in software engineering.

---

[29] See source code at http://groups.umd.umich.edu/cis/course.des/cis400/ada/hworld.html

[30] Compare Christopher McDonald, "From Art Form to Engineering Discipline? A History of US Military Software Development Standards, 1974-1998," *IEEE Annals of the History of Computing* 32, no. 4 (2010): 32-47 at doi.org/10.1109/MAHC.2009.58 with Christopher Tozzi, *For Fun and Profit: A History of the Free and Open Source Software Revolution* (Cambridge: MIT Press, 2017)

[31] Wirth taught at Stanford University (1963-67), was professor of informatics at ETH-Zurich (beginning 1968) and then founding professor in the ETH's new Computer Science department (1981-98); see Dag Spicer, "Niklaus Wirth Obituary," *IEEE Annals of the History of Computing* 46, no. 1 (2024): 74-74 at doi.org/10.1109/MAHC.2024.3366628 .

Since 2018, a series of workshops on gender equality/equity in software engineering—embedded in and co-located with the International Conference on Software Engineering, or ICSE—aims at bringing critical and empirical assessment of gender into the mainstream of software engineering.  ICSE is the longest-running of seven leading software engineering conferences, and it makes an effort to cover the profession comprehensively while other more-specialized conferences deal with testing, models, metrics, and other dimensions of software engineering or with specific application domains.[33]  ICSE has been jointly sponsored by the ACM and SIGSOFT, its special-interest group for software, as well as by the IEEE's Computer Society and the IEEE technical council on software engineering.  In 2018 the ICSE met in Gothenburg, Sweden, with supporters including companies like Facebook, Ericsson, Microsoft, Amazon Web Services, Google, Huawei, and IBM Research, several funding agencies including NSF, and a bevy of prominent Swedish universities and research institutes.

Introducing the first International Workshop on Gender Equality in Software Engineering in 2018, Erika Abraham, Elisabetta Di Nitto, and Raffaela Mirandola positioned it in the wider concern for gender in computing as well as to identify specific concerns in software engineering.  They noted, "diversity [in gender, culture, religion, and geography] plays a key role to a successful and competitive context for software development and research."[34] The five workshops have examined many pertinent topics, such as bias in machine learning and artificial intelligence, intersectional analysis, gender-identifying software, career changing, hackathons, and gender theory. Rather than survey all 53 workshop papers (2018–2024), the following review will focus on gender dimensions of software engineering conferences—to frame the present paper's ICSE data.

---

[32] An active critical field is evident in Thomas S. Mullaney, Benjamin Peters, Mar Hicks and Kavita Philip, eds., *Your Computer Is on Fire* (Cambridge: MIT Press, 2021); Meredith Broussard, *More than a Glitch: Confronting Race, Gender, and Ability Bias in Tech* (Cambridge: MIT Press, 2023); and Jeffrey R. Yost and Gerardo Con Díaz, eds., *Just Code: Power, Inequality, and the Political Economy of IT* (Baltimore: Johns Hopkins University Press, 2025).

[33] Jeffrey C. Carver and Alexander Serebrenik, "Gender in Software Engineering," *IEEE Software* 36, no. 6 (2019): 76-78 at doi.org/10.1109/MS.2019.2934584 .  According to Tao Xie's extensive website on "Software Engineering Conferences," the seven top general SE conferences are ICSE, FSE/ESEC, ASE, SPLASH/OOPSLA, ECOOP, ISSTA, and FASE; see https://web.archive.org/web/20240417212217/https://taoxie.cs.illinois.edu/seconferences.htm

[34] "Message from the GE 2018 Chairs," *IEEE/ACM 1st International Workshop on Gender Equality in Software Engineering* (Gothenburg, Sweden, 2018): quote p. x at https://ieeexplore.ieee.org/xpl/conhome/8452040/proceeding

Participation and visibility at conferences is critical for establishing and building professional careers. Lori Clarke (noted below for leadership in ICSE) and co-authors associated with CRA-W (formally, the Computing Research Association's Committee on the Status of Women in Computing Research) accented the positive role that discipline-specific conferences can play in assisting "underrepresented groups to network and build a community within their subfield." Since the 1990s CRA-W has facilitated such community building, hands-on networking, data-rich surveying, and active mentoring across undergraduate through early-career stages. With women's participation in computing edging up slightly, the CRA-W members were guardedly "optimistic that the landscape is changing."[35]

Two papers analyzed women in conference leadership. The roles of general chair, program chair, program committee, and invited keynote lecturers are clear signs of visibility. For software testing, a branch of software engineering, women overall varied between 17% and 20% of the leadership positions as conference chairs, keynote speakers, and program-committee members (based on 72 conferences in the single year 2016). But one striking difference emerged: in the 40 academic software-testing conferences, just 12.8% of the 2016 keynote speakers were women, whereas in the 32 industrial ones fully 27.8% were women.[36] Academic-industry differences in the roles of conference chair/organizer and program committee were not so pronounced; the proportion of women organizers across all three roles was 18%, varying slightly between academic (17%) and industry conferences (22%).

---

[35] Lori Clarke, Lori Pollock, Jane Stout, Carla Ellis, Tracy Camp, Betsy Bizot, and Kathryn S. McKinley, "Improving Diversity in Computing Research: An Overview of CRA-W Activities," In *IEEE/ACM 1st International Workshop on Gender Equality in Software Engineering* (Gothenburg, Sweden, 2018): 41-44, quotes pp. 43, 44, at doi.org/10.1145/3195570.3195577

[36] "Academic conferences … are usually organized by a university or other research institute, and are often supported by organizations like ACM or IEEE. On the other hand, industrial conferences are more practitioner oriented, and are typically sponsored by companies or other for-profit or non-profit organizations." Judit Jász and Árpád Beszédes, "Software Testing Conferences and Women." Quoted in *IEEE/ACM 1st International Workshop on Gender Equality in Software Engineering* (May 28, 2018, Gothenburg, Sweden): 17-20, quote p. 18, at doi.org/10.1145/3195570.3195582 . The authors note, "Extending the experiments would be interesting … such as by looking at conference attendees (if data is available)."

A second paper investigated women in conference leadership based on six highly ranked software engineering conferences from 2008 to 2017.[37] The authors sought to identify gender disparities in the key roles of general chair, program chair, invited keynote speakers, and program committee members for the main research track. For IEEE conferences, the general chair and program chair are appointed by the conference's steering committee, itself composed of past general and program chairs. Women's participation in these key leadership roles is significantly lower than the percentage of women (37) submitting papers to the 2017 ICSE. Despite hopes of progress, the "females in these visible [conference] roles [have] constantly remained low in the last ten years," the authors report. They also found, distressingly, little correlation between conferences having a female general chair or program chair with the participation of women in other visible conference roles. Clearly, here as elsewhere, greater efforts from the leadership in computing are needed.

Since roughly the 2000s, the computer science community has invested tremendous resources addressing gender disparities. Some modest success is reported in the most recent (2024) Gender Equity workshop.[38] Moldovan and Motogna focused on diversity measures of the program committees for three leading software engineering conferences, including ICSE, during the years 2019 to 2023, analyzing gender, geography, seniority of rank, industry/academic, as well as new program committee members (an aim is to have one-third new members each year as well as at least 10 percent from industry). Notably, unlike many recent studies relying on gender-ID software such as NamSor, "gender and geographical location [for] each PC member [was] manually performed." The authors identified differences among the three conferences, with ASE (IEEE/ACM International Conference on Automated Software Engineering) the least female friendly with 24.3 percent of program committee members being female; the ESEC/FSE (ACM Joint European Software Engineering Conference and Symposium on the Foundations of Software Engineering) in the middle with 33.6 percent female; and the top-ranked ICSE the most female friendly at 37.6 percent. (Non-binary program committee members constituted 0.4, 1.2, and 0.7 percent, respectively.) The authors noted some success in that Europe and North America —across all three conferences—had relatively smaller differences (than other regions such as Asia, South America, and Australia, while overall "Africa is almost missing") in male and female program-committee members, "an improvement in gender diversity." The authors note that ICSE achieved this laudable gender diversity because it is a "highly prestigious conference," with stringent acceptance, sizable number of accepted papers, and large number of participants and co-located events (such as the Gender Equity workshops) and so able to easily draw on a larger and more diverse pool to create its program committees.[39]

---

[37] The six conferences were, in addition to ICSE, ASE, FSE, RA, EASE, and ESEM (compare Tao Xie supra). Muneera Bano and Didar Zowghi, "Gender Disparity in the Governance of Software Engineering Conferences." *IEEE/ACM 2nd International Workshop on Gender Equality in Software Engineering* (Montreal, Canada, 2019): 21-24, quote p. 22, at doi.org/10.1109/GE.2019.00016.

[38] Vasilica Moldovan and Simona Motogna, "Diversity of SE Conferences," in *ACM/IEEE Workshop on Gender Equality, Diversity, and Inclusion in Software Engineering* (New York: ACM, 2024): 47-54 at doi.org/10.1145/3643785.3648492 . Compare Aditya Shankar Narayanan, Dheeraj Vagavolu, Nancy A. Day, and Meiyappan Nagappan, "Diversity in Software Engineering Conferences and Journals," *ArXiv* (2023): 13 pages at doi.org/10.48550/arXiv.2310.16132 which (based on three leading software-engineering conferences and two top journals in the field) found surprisingly little correlation between diversity of program committee and diversity of authors on the resulting program; and, furthermore, while there was little improvement in gender or geographical diversity, ethnic diversity improved during 2010-22.

**Measuring gender bias**

Women expanded their participation as research authors in ICSE across the period 1976–2010, rising from under ten percent to substantially over 20 percent (see figure 1). Such growth in women's participation in computer-science research is consistent with longitudinal data from the National Science Foundation, the US Bureau of Labor Statistics, as well at the author's published findings based on the DBPL dataset, ACM publications, and computing user groups.[40] These findings do not lend much credence to the "making programming masculine" conjecture that has become popular in the history of computing community and beyond.[41] While attractive as a sociological theory about professionalization in computing, there is little empirical evidence that women left the computing workforce or the computer-science research community during the 1960s and 1970s in the decades of active professionalization, as the "making programming masculine" conjecture posits. Women's participation in computing actually expanded throughout the 1960s and 1970s and into the 1980s. In the mid-1980s, women in the US achieved a peak in computer-science undergraduate degrees and in the high-skilled computing workforce at, respectively, 37 percent and 38 percent. Women's proportion of computing, while recovering modestly in the past decade or so, remains (through today) substantially lower than in the mid-1980s. It is abundantly clear that women's proportion of computing jobs and share of computing research is worrisomely low today;[42] it is also inescapable that this trend has its origins in the mid-1980s.

---

[39] This empirical finding does not support the hypothesis (embedded in the "making programing masculine" conjecture) that higher prestige sub-fields or status-seeking activities, like professionalization, intentionally increase gender exclusivity in order to reduce (supposedly lower-status) female participation. Acceptance rates for conferences are a commonly used index of prestige in computer science: ICSE acceptance rate 16.9%; ASE, 17.9%; ESEC/FSE, 22.6%.

[40] "Dynamics of Gender Bias in Computing," *Communications of the ACM* 64 no. 6 (June 2021): 76-83 at doi.org/10.1145/3417517 ; "Gender Bias in Big Data Analysis," *Information and Culture* 57 no. 3 (2022): 283-306 at doi.org/10.7560/IC57303; "Dynamics of Gender Bias within Computer Science," *Information and Culture* 59 no. 2 (2024): 161-81 at doi.org/10.7560/IC59203

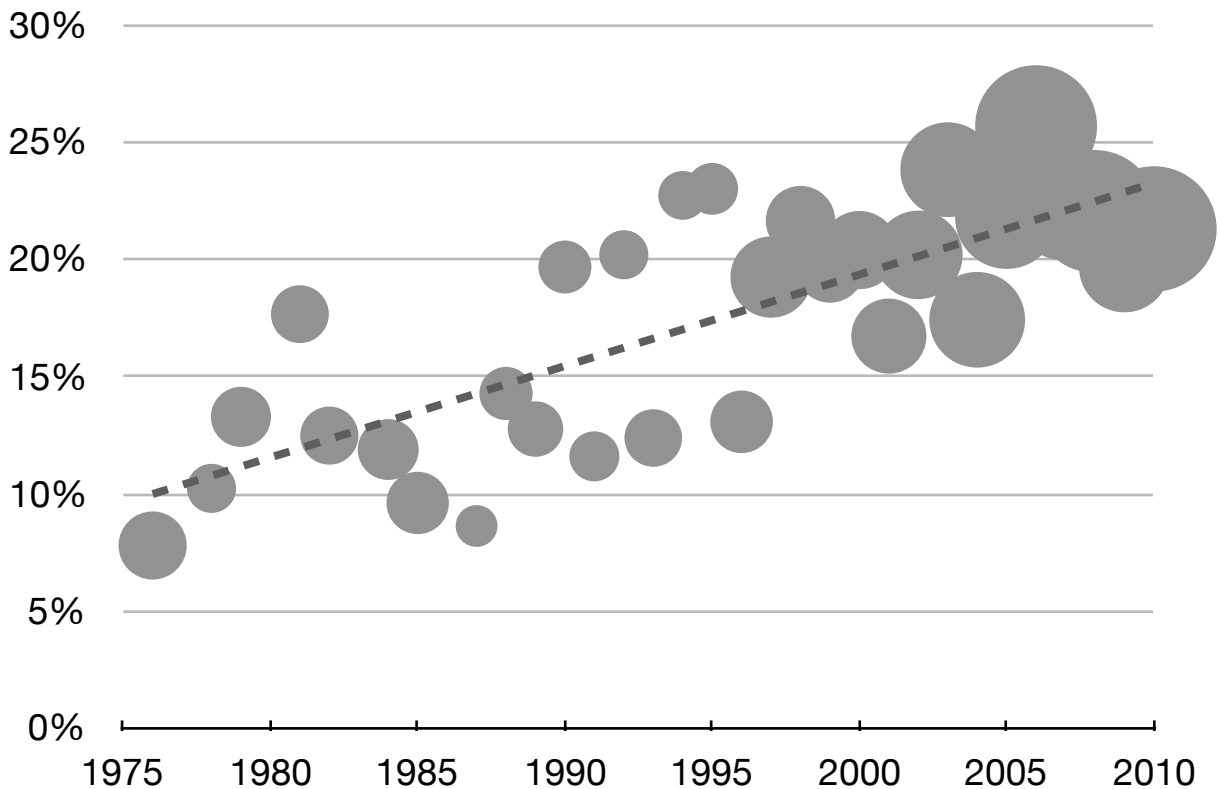

Figure 1: Women as research authors in ICSE (1976–2010)

---

[41] "Computer programming was originally a women's discipline, seen as lowly and technical work. Over a short time, . . . the gender representation in the discipline changed to male-dominated, and with it the esteem and pay for the discipline," writes Stephen Secules, "Making the Familiar Strange: An Ethnographic Scholarship of Integration Contextualizing Engineering Educational Culture as Masculine and Competitive," *Engineering Studies* 11 no. 3 (2019): 196-216, citing (on p. 209) Nathan Ensmenger's "Making Programming Masculine," *Gender Codes: Why Women Are Leaving Computing* (2010), 115-141 at doi.org/10.1002/9780470619926.ch6. Also citing Ensmenger's thesis are Nikolay Rudenko, Irina Antoshchuk, Roman Maliushkin and Liliia Zemnukhova, "Gender Equality Paradise Revisited: The Dynamics of Gender Disbalance in Russian Engineering from the Late Soviet Time to the 2010s," *Engineering Studies* 14 no. 1 (2022): 56-78, on p. 67; and Kristin A. Bartlett and Stephanie M. Masta, "A Glimpse into the Gendered Dynamics in Industrial Design through the Podcast Discourse," *Engineering Studies* 15 no. 3 (2023): 180-200, on p. 194. An assessment is provided in William F. Vogel, "'The Spitting Image of a Woman Programmer': Changing Portrayals of Women in the American Computing Industry, 1958-1985," *IEEE Annals of the History of Computing* 39 no. 2 (2017): 49-64 at doi.org/10.1109/MAHC.2017.14.

Just what *changed* about computing beginning in the 1980s needs additional and focused analysis. One Gender Equity workshop paper has suggested that computer science departments, often in engineering schools, imposed restrictive recruitment and curricular practices in an attempt to handle the 1980s "boom" in computer science enrollments that overwhelmed their capacities. For example, Purdue imposed new restrictions on students entering the computer-science major, slashing overall CS enrollments in just a few years, while the University of Maryland (another leading department) also imposed entry restrictions like GPA requirements and additionally strictly limited class sizes. Other steps taken by some CS departments included "exclusionary policy actions," most notably "the retooling of first-year CS as a weeder course [resulting] in a competitive, 'chilly' atmosphere that disproportionately discouraged women." With the dot-com boom in the 1990s and then bust around 2000, another another cycle of gender exclusionary practices took form as many computer-science departments expanded or entrenched their engineering-specific profile (e.g. emphasizing engineering to strengthen their cases for additional faculty hiring and/or creating "computer engineering"[43]) while downplaying the importance of so-called soft skills and application to real-world problems. These gender exclusionary changes were reinforced by the open-source movement that was sometimes flagrantly hostile to women.[44] The ICSE data presented below offer a striking empirical case of gender bias in the 1980s.

---

[42] The journal *IEEE Software* has twice declared a "diversity crisis": Bram Adams and Foutse Khomh, "The Diversity Crisis of Software Engineering for Artificial Intelligence," *IEEE Software* 37, no. 5 (2020): 104-108 at doi.org/10.1109/MS.2020.2975075 ; Khaled Albusays, Pernille Bjorn, Laura Dabbish, Denae Ford, Emerson Murphy-Hill, Alexander Serebrenik, Margaret-Anne Storey, "The Diversity Crisis in Software Development," *IEEE Software* 38, no. 2 (2021): 19-25 at doi.org/10.1109/MS.2020.3045817.

[43] Rachel E. Friedensen, Sarah Rodriguez and Erin Doran, "The Making of 'Ideal' Electrical and Computer Engineers: A Departmental Document Analysis," *Engineering Studies* 12 no. 2 (2020): 104-126; Brent K. Jesiek, "The Origins and Early History of Computer Engineering in the United States," *IEEE Annals of the History of Computing* 35, no. 3 (2013): 6-18 at doi.org/10.1109/MAHC.2013.2.

The bibliometric data in this paper results from analysis of research papers published in the ICSE conference series beginning in 1976 through 2010. These ICSE papers (n= 2,652) were downloaded from the IEEE Digital Library (February 2024), names of authors and coauthors were extracted, instances of multiple authorship in each year were combined: the result comprises all research authors from that year's conference, creating a comprehensive dataset (n = 6,102) to assess women's research participation in software engineering.[45] Next, a historically sensitive gender analysis of authors' first names, adjusted for the year of publication, was performed. Historical research using commonly available gender-identification software (such as Gender-API, NamSor, and Genderizer.io) often gives inaccurate results, since these software tools' reliance on present-day name–gender associations is inaccurate for earlier years when these associations can be dramatically different. The "Leslie problem" illustrates the point: Leslie in 1900 was firmly a male name, used 92 percent of the time for naming boys; around 1950, Leslie was used equally for naming girls and boys (respectively, 52 and 48 percent); while since around 2000 it is preponderantly female (95 percent for girls). Using gender-ID software that is effectively trained on today's use of Leslie as a female name obviously undercounts the male Leslies of earlier years.  Other names with shifting gender associations include Addison, Allison, Jan, Jean, Kendall, Madison, Morgan, and Sydney (a separate publication assesses 300 US names with significant gender shifts from 1925 to 1975).  An overall net "female shift" in names during these decades means that, on balance, these software tools will *undercount* men with gender-shifting names and consequentially lead to an *overestimate* of women in the early years of computing.[46]

---

[44] Elizabeth Patitsas, "The Social Closure of Undergraduate Computing: Lessons For The Contemporary Enrolment [sic] Boom," *IEEE/ACM 2nd International Workshop on Gender Equality in Software Engineering* (Montreal, Canada, 2019): 33-36 at doi.org/10.1109/GE.2019.00015 ; Vandana Singh and William Brandon, "Discrimination, Misogyny and Harassment: Examples From OSS: Content Analysis of Women-Focused Online Discussion Forums," *Proceedings of the Third Workshop on Gender Equality, Diversity, and Inclusion in Software Engineering* (New York: ACM, 2022): 71-79 at doi.org/10.1145/3524501.3527602; Hana Frluckaj, Laura Dabbish, David Gray Widder, Huilian Sophie Qiu, and James D. Herbsleb, "Gender and Participation in Open Source Software Development," *Proceedings of the ACM on Human-Computer Interaction*, Volume 6, Issue CSCW2, No. 299 (2022): 1-31 at doi.org/10.1145/3555190

[45] The position of author's names (especially first and/or last) can in some fields reveal significant contributions to a multi-author paper. There does not seem to be such consensus in computer science. For an exploratory study (contrasting the subfields of machine learning and human-computer interaction) see Kirstin Early, Jessica Hammer, Megan Kelly Hofmann, Jennifer A. Rode, Anna Wong, and Jennifer Mankoff, "Understanding Gender Equity in Author Order Assignment," *Proceedings of the ACM on Human-Computer Interaction*, Volume 2, CSCW, no. 46, (2018): 1-21 at doi.org/10.1145/3274315. Findings (page 6) are that "author order can be determined by rules, negotiation, and randomness" with interviewees offering such comments as: "really/very complicated," "thorny," "not always easy to determine," and "really ad hoc."

    Careful use of the Social Security Administration (SSA) dataset of all instances of each name used in the United States since the 1880s offers an alternative.  At least since the 1940s, when the SSA corrected its undercounting of agricultural workers, the SSA dataset offers an immense sample of US names.  Furthermore, the SSA data can be used to "tune" name–gender associations to be historically accurate.  Ideally, one might look up the SSA results for the author's year of birth: in 1980 "Jamie" was used 2,846 times for naming baby boys, and 11,526 times for baby girls, so the probability of an average Jamie born that year being female is 0.802.  Author's birthdates, of course, are rarely known.  My earlier research with (manually) gender-identified authors indicated that optimum results could be achieved with a 'year shift' of 30 (that is, doing an author's first-name look-up in the SSA dataset 30 years prior to the author's research publication).  One naturally expects with the expansion of computer science in Asia—and accordingly more authors with East Asian and South Asian names that might not be in the SSA dataset—that the proportion of ICSE author names *not* found in the US-based SSA data could be increasing.  But the actual median of non-SSA names in the last study-decade (2000-10) of 25.0

---

[46] Relying on the Gender-API tool, Lucy Wang and colleagues reported that women composed 20 percent of research authors in computer science in the early 1950s; see Lucy Lu Wang, Gabriel Stanovsky, Luca Weihs, and Oren Etzioni, "Gender Trends in Computer Science Authorship," *Communications of the ACM* 64 no. 3 (March 2021): 83 figure 4 at doi.org/10.1145/3430803.  This estimate is fully ten times above previous published sources: women were 1.8 percent of early CS authors (1949-54) according to José María Cavero et al., "The Evolution of Female Authorship in Computing Research," *Scientometrics* 103 (2015): 89 at doi.org/10.1007/s11192-014-1520-3; the present author (separate article) found 1.7 percent women authors in the DBPL dataset (1950-54). Women earned fewer than 3 percent of US doctoral degrees in computer science during 1966-70 according to J. McGrath Cohoon and William Aspray, eds., *Women and Information Technology: Research on Underrepresentation* (Cambridge: MIT Press, 2006), p. x.  Wang et al.'s elevated findings also for women in engineering and medical research in the 1950s suggests an underlying problem with their analysis.

percent is just slightly above the median figure across the study years (1976-2010) of 23.3 percent.[47]

Analysis of ICSE conferences permits fine-grained analysis of the software engineering community. From 1976 onward, ICSE took its international mandate seriously, with conferences regularly meeting outside North America; since 1987 ICSE has met annually with the same international emphasis. ICSE meetings in Japan in 1982 and 1999 had the greatest number of non-SSA names as authors, each over 40 percent; the 2006 meeting in China was third highest at 37.4 percent. Year-by-year analysis of such international meetings also can pinpoint a striking pattern of gender bias in software engineering in the 1980s and early 1990s.

There are some anomalies with figure 1. Overall while the trend is up, only modest correlation exists between the computed linear trendline and actual data points. On inspection, there is highly unusual meeting-by-meeting variance from 1982 through 1994, when women's proportion as ICSE authors jumped around between 9 percent and 22 percent. Furthermore, the international meetings and the Anglo-American meetings in these years were systematically different for women. One might conjecture that the 1982-94 variation could indicate that women experienced greater difficulties in securing travel to the international meeting sites: Singapore, Tokyo (Japan), Melbourne (Australia), Nice (France), and Sorrento (Italy). But, in fact, it was not "international" meetings during these years where women experienced difficulties in presenting their research, networking, and developing their careers; rather, the problem was the US meetings during these years, including London in 1985.

The pattern of women experiencing greater barriers during these years in US and UK meetings is inescapable (see figure 2). At ICSE meetings held in Monterey CA, Baltimore MD, Austin TX, Pittsburgh PA, Orlando FL, and London women's participation ranged from 9.63 percent to 12.76 percent (median = 11.74 percent). Conversely, at ICSE meetings held at the five "international" locations, noted above, their participation ranged from 12.48 percent to 22.72

---

[47] Computer science researchers have developed neural-network methods to assess the gender of Chinese names; but it is unclear these methods are valid for historical research (as noted above). See Zihao Pan, Kai Peng, Shuai Ling, and Haipeng Zhang, "For the Underrepresented in Gender Bias Research: Chinese Name Gender Prediction with Heterogeneous Graph Attention Network," *Proceedings of the AAAI Conference on Artificial Intelligence* 37 no.12 (2023): 14436-14443 at doi.org/10.1609/aaai.v37i12.26688

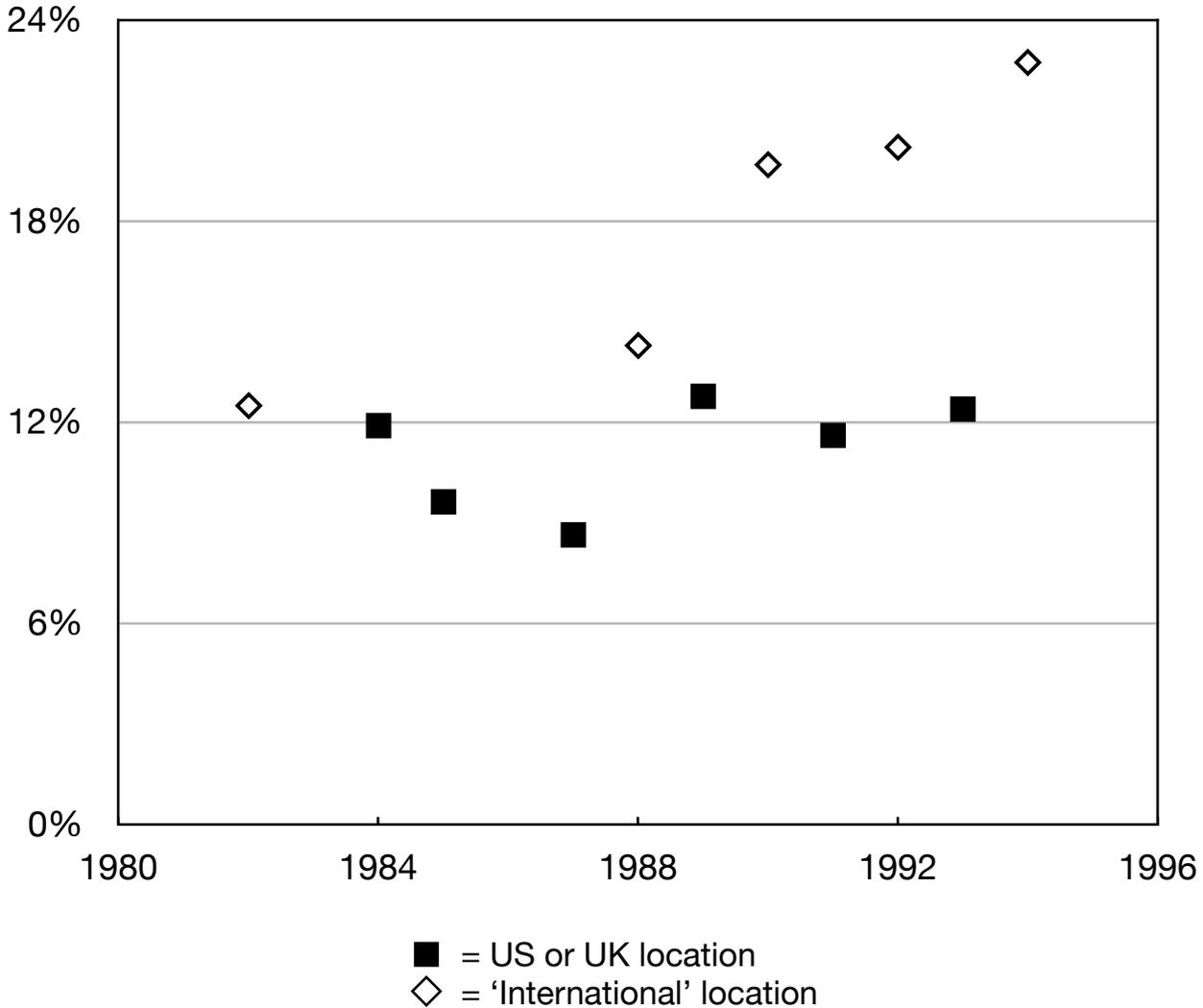

Figure 2: Women as research authors in ICSE (1982-94)

■ = US or UK location
◇ = 'International' location

percent (median = 19.67 percent). It may be tempting to ask whether these differences are statistically "significant" but a careful analysis is needed: a standard statistical test of differences between means, while easily computed, is of dubious statistical value with such small populations (commonly used statistical techniques as the T test require large populations and also 'normal' distributions).

A statistically robust alternative better suited to small populations as here is the Mann-Whitney U test. It is conducted by rank ordering the dataset rather than comparing population mean and standard-deviation values (as used in the common t-test). Since women's participation as ICSE research authors across 1982-94 was increasing overall, we can examine how much "above" or "below" each meeting might have been assuming a simple linear increase across these years.  The US meetings were uniformly *below* their expected value (ranging from 1.0 to 5.7 percent below the expected linear value); the London meeting was 3.8 percent below.  The international meetings were all at or *above* their expected value (ranging up to 5.2 percent for the Sorrento, Italy, meeting). The Mann Whitney U test, computed for both the 'raw' and 'adjusted' data, was found to be statistically significant at p < .01.[48]  That is to say: these differences— women's participation at the international versus the Anglo-US meetings—are unlikely (with a probability of less than one in 100) to have resulted from chance alone.

      The most obvious point is plain: in these years women experienced significant barriers to participation at ICSE conferences held in the US and UK.  Let's consider some possibilities.  The Gender Equity papers (reviewed above) direct attention to the key roles of women in conference leadership.  From 1976 through 2002 the ICSE leadership positions of general chair and program chair(s) were entirely dominated by men, so the statistically significant gender disparity during 1982-94 was (of course) not the result of men's versus women's leadership.  Interestingly,  the first meeting with women in ICSE leadership occurred in 2003 in Portland, Oregon; that year Lori Clarke served as general conference chair while Laurie Dillon and Walter Tichy served as program chairs.  Notably, that year women's percentage as research authors jumped from a baseline of 16.7 and 20.2 in the two years prior, up to 23.8 in 2003, then dropped back to 17.4 and 21.8 percent in the two years following.  (This is strong positive evidence that, for 2003 at least, women's conference leadership resulted in expanded women's conference participation.) Beginning with 2003, women regularly served as general and/or program chairs; in addition to Lori Clarke and Laurie Dillon, other women in these leadership roles included Mary Lou Soffa, Joanne M. Atlee, Paola Inverardi, and Judith Bishop.  From 2003 to 2010, these six women

---

    [48] See Michael P. Fay and Michael A. Proschan, "Wilcoxon–Mann–Whitney or t-test? On assumptions for hypothesis tests and multiple interpretations of decision rules," *Statistics Surveys* 4 (2010): 1-39 at doi.org/10.1214/09-SS051 or W. J. Conover's *Practical Nonparametric Statistics* (New York: John Wiley & Sons, 1999).  Consulting socscistatistics.com for a Mann-Whitney U test using the raw data (actual percentages of women's research authorship): The U-value is 1. The critical value of U at p<.01 is 2. Therefore, the result is significant at p < .01.  For the year-adjusted (expected) data, the U-value was 0, leading again to a finding of significance at p < .01. Associated Z-scores, typically considered valuable when N > 10, are not here reported (but were also significant at p < .01)

constituted 24 percent of the ICSE conference leadership.  The meetings with women's leadership spanned the globe: in addition to Portland, Oregon, there were also Shanghai, China (2006), Vancouver, British Columbia (2009), and Cape Town, South Africa (2010). One can posit significant differences in women's ICSE participation during years with—and without—women in leadership roles; on examination, however, the variations during the years 2003-10 were modest and the results — unlike 1982-94 — do not appear to be statistically significant.

These dozen years of statistically significant barriers to women's participation in ICSE coincided with unsettling changes in the United States.  Women's impressive decades-long expansion in undergraduate computer-science degrees, and participation in the high-skilled US computing workforce, reversed and began falling dramatically.  During these same years, the all-male ICSE conference leadership—when ICSE was meeting in the US or UK—created ICSE conference programs with significantly reduced women's participation.  The existing documentation on ICSE's internal decision making (i.e., how and why conference-paper proposals were accepted) is thin and does not readily permit inference of possible motives for this clear instance of gender exclusion.

An ICSE keynote address in 2003, when Lori Clarke chaired the conference, suggests that ICSE knew there was a gender problem and that Clarke as general chair saw the need for addressing it.[49]  Gender scholar Joanne M. Cohoon's "Must There Be So Few? Including Women in CS" adroitly assembled the worrisome empirical evidence (there is little on software engineering *per se* in its published version).  Cohoon held faculty positions at the University of Virginia and also became senior researcher at the National Center for Women and Information Technology, founded in 2004.  She framed her ICSE keynote with comparative data from 1970–1999 that demonstrated while women's percentage of computer-science undergraduate degrees expanded to a peak in the mid-1980s, undergraduate women in the fields of biology/life sciences and mathematics expanded too; undergraduate women students in physical science grew steadily to exceed those in computer science, with traditionally male-heavy engineering almost reaching computer science (with broadly parallel shifts in computer-science master's and doctoral degrees).[50] She next indicated a worrisome drop that occurred from 1991 to 2001 in college-entering women intending to major in computer science: while 37 percent of computer-science-intending majors had been women in 1991 (in accord with the mid-1980s peak), this dropped to just 20 percent by 2001. Cohoon then presented compelling evidence from her 2000 University of Virginia dissertation, "Non-Parallel Processing: Gendered Attrition in Academic Computer Science," which documented significant differences in male students' and female students' *annual attrition* rates in computer science undergraduate degree programs (15 and 21 percent, respectively); women, already fewer in computer-science programs, left these programs at significantly higher rates. Cohoon did highlight Carnegie Mellon University, which actually boosted its women CS majors from 7 to 40 percent from 1995 to 2000.[51] Cohoon's keynote was the very first time "women" appeared anywhere in the title of ICSE papers.

---

[49] By 2003, the drop of women in computing was public news: Karen Stabiner, "Where the Girls Aren't," *New York Times* (12 January 2003) at https://www.nytimes.com/2003/01/12/education/where-the-girls-aren-t.html (accessed January 2025)

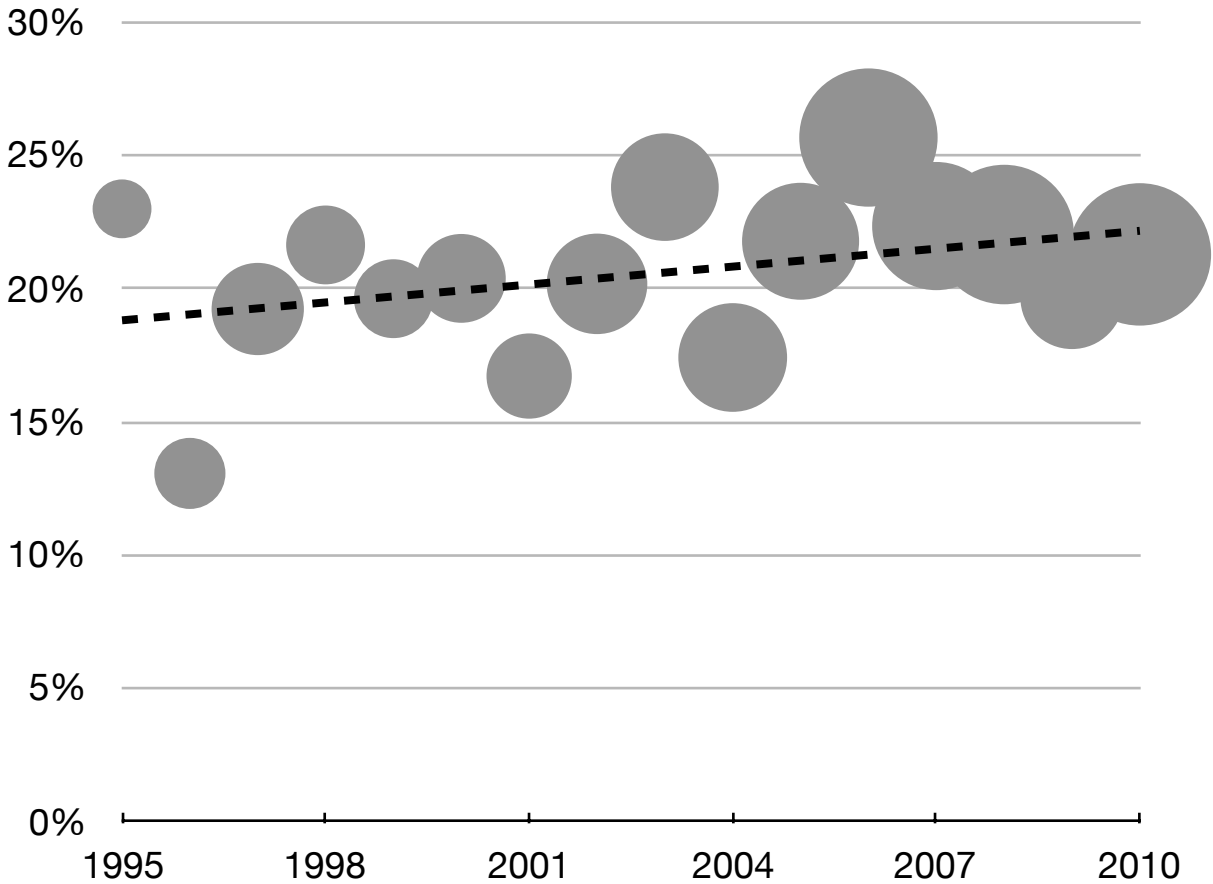

Figure 3: Women as research authors in ICSE (1995–2010)

---

[50] According to NSF, in 2008 engineering overtook computer science in its percentage of undergraduate degrees awarded to women, with 18.5 and 17.7 percent, respectively; see National Science Board, *Science & Engineering Indicators 2018* (Alexandria, VA: National Science Foundation, 2018), figure 2-11 at https://www.nsf.gov/statistics/2018/nsb20181/figures#chapter561

Figure 3 shows women in ICSE 1995-2010, after the blatant gender exclusionary dozen years had passed. There is quite a different trend from figures 1 or 2. In these recent 15 years there was much slower growth in women's participation, around half the rate from 1976–2010 (the decadal growth fell from 3.9 percent to 2.2 percent). Put another way, if the growth rate across this study period's 35 years persisted, women in software engineering might achieve rough parity with men by 2040; with the recent slower growth rate during 1995-2010, women would achieve rough parity only by 2070.

## Conclusions

This longitudinal analysis of women's participation in software engineering both challenges and extends findings about the nature of engineering as well as computer science. While patterns of gender bias are entrenched and pernicious, attention to historical changes in these patterns is often overlooked. Gender bias in computing did not "switch on" sometime in the 1950s or 1960s or 1970s; instead, computing during these decades was relatively hospitable to women, compared with engineering and physical science. Beginning in the 1980s, software engineering experienced a dozen years when women faced (statistically) significant barriers to their participation in the ICSE community: during these years, international meetings were less gender biased than the US and UK meetings. These same years witnessed the onset of today's pattern of gender bias, with women's participation falling to roughly half the mid-1980s peak. Software engineering appears an instance of early onset to the reversal of women's decades-long growth in computing from the 1960s to the mid-1980s.

These findings can also be considered "policy relevant" in the following way. While much academic research concerns itself with debates and contributions within a disciplinary or multidisciplinary structure, the research on gender in computing naturally connects with the numerous policy actors—including the National Science Foundation, Sloan Foundation, National Center for Women & Information Technology, branches of the IEEE and ACM, and, not least, the annual Grace Hopper Celebration of Women in Computing, among others—that are actively striving to improve women's participation in computing, including software engineering.  The policy questions include: What went wrong in computing, and when, and what can can we change today?  Research that is aware of significant alterations and turning points, with women's participation changing substantially across time is one small step.  Another is a long-overdue recognition of the sizable diversity that exists *within* the numerous computing fields.[52]  Here a focus on software engineering spotlights the field's persistent concern with achieving the promise and possibly also the status of "engineering," with an active (if stylized) understanding of engineering history.  Attention to to the dynamic fine structures of the technical professions — to the micro-dynamics if you will — may serve to align research in engineering studies with these pressing policy concerns and society needs.

---

[51] See Jane Margolis and Allan Fisher, *Unlocking the Clubhouse: Women in Computing* (Cambridge: MIT Press, 2001).

---

[52] See the author's "Dynamics of Gender Bias within Computer Science," *Information and Culture* 59 no. 2 (2024): 161-81.